\documentclass[preprint,12pt,authoryear=false]{elsarticle}
\usepackage[T1]{fontenc}
\usepackage[utf8]{inputenc}
\usepackage{lmodern}
\usepackage{graphicx,booktabs,longtable,array,amsmath,amssymb,float}
\usepackage{microtype}
\usepackage{url}
\usepackage{placeins}
\usepackage[breaklinks=true,hidelinks]{hyperref}
\journal{International Journal of Critical Infrastructure Protection}
\graphicspath{{}}
\begin{document}
\begin{frontmatter}
\title{Compound Environmental Operating Stress and Cyber-Induced Pump Disruption in Water Distribution Networks: Matched Hydraulic Impacts, Service Continuity, and Sustained Recovery}
\author[wmu]{Halil Ibrahim Dursunoglu\corref{cor1}}
\ead{halilibrahim.dursunoglu@wmich.edu}
\author[gov]{Fatma Dursunoglu}
\ead{fatmaulkukaratass@gmail.com}
\cortext[cor1]{Corresponding author.}
\affiliation[wmu]{organization={Department of Computer Science, Western Michigan University},city={Kalamazoo},state={Michigan},postcode={49008},country={USA}}
\affiliation[gov]{organization={General Directorate of Combating Desertification and Erosion, Ministry of Environment, Urbanization and Climate Change},city={Ankara},postcode={06560},country={T\"urkiye}}
\begin{abstract}
Digitally controlled water distribution systems must maintain
service during disruptions, yet restoring water delivery does
not necessarily ensure hydraulic recovery. This study evaluates
cyber-induced pump outages under hypothetical environmental
operating stress using two EPANET benchmark networks (Net1
and Net3) simulated with the Water Network Tool for Resilience.
The experimental design comprises 288 simulation records
(240 distinct configurations), including 192 verified attack
scenarios, with four- and twelve-hour outages, six hydraulic
configurations, and four mitigation strategies. All attack
scenarios satisfied pump-shutdown verification and pre-attack
hydraulic-equivalence criteria. Twelve-hour outages produced
an average of 404.96 m$^3$ of unmet demand in Net1, while
Net3 maintained full modeled demand delivery despite pressure
deviations. Sustained pressure recovery occurred in 103 of
192 scenarios, with 65 exhibiting subsequent relapse, whereas
all scenarios satisfied the post-outage water-service recovery
criterion. Combined mitigation eliminated unmet demand in
evaluated Net1 scenarios, while outage shortening achieved
the highest sustained-pressure-recovery count. These findings
highlight network-dependent resilience and the distinction
between service restoration and hydraulic recovery. Environmental
stress scenarios represent sensitivity assumptions rather
than empirically calibrated climate impacts.
\end{abstract}
\end{frontmatter}
\section{Introduction}\label{introduction}

Classical water-system performance research distinguishes reliability, recovery speed, and failure severity \cite{hashimoto}, while network resilience indices emphasize the available hydraulic surplus and network design \cite{todini}. Water distribution systems integrate pumps, storage tanks, reservoirs, pipes, and increasingly automated controls. This cyber--physical dependence enables efficient operations but introduces pathways by which unauthorized operational changes may affect service delivery. NIST's guidance for operational technology stresses that cybersecurity decisions must consider physical reliability and safety \cite{nist}. The EPANET 2.2 hydraulic solver provides an established platform for extended-period distribution simulations \cite{epanet}. Modeling work on water systems has advanced from anomaly detection to attack-aware control and co-simulation \cite{klise,taormina,housh,shin,karrenberg}. Yet a practically important distinction remains: after an attack stops, a network may again satisfy requested demand while still differing from the hydraulic trajectory it would have followed without the attack.

Resilience measurements frequently emphasize continuity or rapid restoration of a selected service metric. In pressurized water systems, however, storage and feedback controls introduce state dependence. An interruption may deplete a tank, cause controller switching, and change subsequent pressure trajectories. A temporary return within an acceptable pressure range is not necessarily stable recovery, because controls and storage states can drive later divergence. Consequently, both the counterfactual comparison and the definition of recovery warrant careful validation.

Hydraulic stress is also relevant to climate adaptation: shifts in demand or available source head can change how a network absorbs pump disruption. In the absence of calibrated utility-specific climate--hydraulic relationships, however, modifying reservoir head or nodal demand is a sensitivity test rather than a measured climate-change effect. This study therefore frames environmental operating stress as a controlled hydraulic factor and does not attribute simulated outcomes to historical drought episodes.

The study asks: (RQ1) How do verified pump outages affect pressure and demand delivery in distinct benchmark networks? (RQ2) Do prescribed demand/head stresses modify marginal attack effects? (RQ3) How do mitigation configurations compare on absolute service loss and sustained recovery? (RQ4) How often do trajectories exhibit apparent recovery followed by relapse? Our contribution is an auditable paired-simulation workflow and a multi-outcome analysis that separates service restoration, first pressure recovery, sustained recovery, and relapse.

\subsection{Motivation from climate-related water stress}\label{motivation-from-climate-related-water-stress}

Climate change provides the long-term environmental motivation for the compound-risk framework. Temperature, precipitation, evapotranspiration, hydrological drought, and seasonal water use may influence the operating circumstances facing a water utility, but their paths to actual pump scheduling, reservoir supply head, and customer demands depend on local infrastructure and policy. Water-supply resilience therefore requires distinction among (i) historical observations at the watershed/county scale; (ii) calibrated supply-system operational responses; and (iii) hypothetical hydraulic operating scenarios. This study rigorously evaluates tier (iii) and makes the evidence needed for tiers (i)--(ii) explicit. In particular, a regional river gauge is not presumed to be the hydraulic supply source for Net1 or Net3.

From a resilience perspective, the question is not whether a drought observation automatically reduces a benchmark reservoir head by a fixed amount. Rather, it is whether a pump-control interruption produces different service and recovery consequences when the system is placed in alternative source-head or demand conditions, and whether those conditions can eventually be linked to empirical environmental drivers. The current analysis answers the former question while preserving the unverified status of the latter.

\subsection{Contributions and research questions}\label{contributions-and-research-questions}

The contributions are fourfold. First, a control-consistency audit requires attacker and reference hydraulics to be identical prior to disturbance. Second, recorded pump status and flow verify that the forced-operational disruption actually occurs in the hydraulic solver. Third, paired analyses quantify service and pressure sensitivity under six hypothetical operating configurations and four defense assumptions. Fourth, longitudinal recovery analysis differentiates first return to an acceptance region, sustained return, and a later relapse. These contributions emphasize model and endpoint validity rather than a new exploit or an empirically proven drought effect.

\section{Related work}\label{related-work}

Network resilience has also been assessed through graph-theoretic sectorization \cite{herrera} and studies connecting network topology to hydraulic performance \cite{meng}. The Safe \& SuRe framework distinguishes reliability, resilience, and sustainability in water-management decisions \cite{butler}. Klise et al.~\cite{klise} introduced WNTR as a resilience modeling framework capable of simulating physical disturbances and recovery actions. Taormina et al.~\cite{taormina} benchmarked cyberattack detection in water distribution networks, while Housh and Ohar \cite{housh} used physically grounded models for detection. Shin et al.~\cite{shin} examined resilience to operational failures arising from cyber--physical attacks, highlighting the relevance of restoration and system interactions. Karrenberg et al.~\cite{karrenberg} applied model checking to cyberattack vulnerabilities and physical consequences. Cohen et al.~\cite{cohen} introduced WaCSim, a co-simulation platform connecting hydraulic modeling with cyber networking and different control architectures, thereby illustrating that hydraulic interruption modeling alone is not novel.

In contrast to claims of a new attack detector or a detailed emulation of actual adversarial access, the present work addresses validity of forced-operational-disruption experiments and the interpretation of recovery trajectories. It does not emulate a complete communication network, prove exploitability of utility controls, estimate attack frequencies, or measure effects in a deployed water utility.

\subsection{Water-network cybersecurity and physical fidelity}\label{water-network-cybersecurity-and-physical-fidelity}

The BATADAL benchmark \cite{taormina} clarified the difficulty of detecting control-system manipulation in water networks, while physically grounded model-based detectors \cite{housh} integrate hydraulic constraints into attack reasoning. Such detection studies pursue questions about whether malicious operations can be recognized in sensor and process data. The present experiment instead applies imposed pump interruptions as exogenous disturbances and asks whether their modeled hydraulic effects and recoveries are computed against defensible counterfactual trajectories. Consequently, high detection accuracy in the prior literature would not substitute for confirming forced pump closure in our testbed.

Hydraulic resilience frameworks such as WNTR \cite{klise} introduced simulation of disrupted operation and post-event service trajectories. Recent cyber--physical research broadens control models and experimental realism. In particular, WaCSim \cite{cohen} couples EPANET-based physical models with emulated industrial communication networks and explicitly evaluates attacks and defenses through control architectures. Our work does not reproduce this richer cyber-layer functionality; it contributes a simpler, auditable pump-outage experiment with explicit sustained-state recovery measurements. The distinction limits claims of novelty while sharpening the methodological contribution.

\subsection{Drought observations, hydrology, and utility hydraulics}\label{drought-observations-hydrology-and-utility-hydraulics}

The U.S. Drought Monitor reports weekly spatial drought classifications, and the U.S. Geological Survey publishes observed river discharge. Both are legitimate environmental measurements. Neither, by itself, measures the operational source elevation or household demand at a specific water utility. Inferring these intermediate states would require local intake configuration, reservoir operation, aquifer or surface-water connection, service-area demand and conservation policies, and seasonally adjusted utility records. The experimental factor \emph{source-head decrement} is thus an intervention on a benchmark-network parameter, not an estimate derived from a local stream-gauge flow deficit.

The empirical extension analyzes Ingham County drought observations, USGS Red Cedar River discharge, and NASA POWER weather for 2020--2025. The observed association is regional hydrology rather than direct calibration of water-distribution operating conditions. The USGS identifies the station and its historical daily data availability \cite{usgs}; the related U.S. Drought Monitor county statistics service is identified in the provided source code \cite{usdm}.

\section{Materials and methods}\label{materials-and-methods}

\subsection{Networks and hydraulic modeling}\label{networks-and-hydraulic-modeling}

Net1 contains 9 junctions, 12 pipes, one pump, one tank, and one reservoir; Net3 contains 92 junctions, 117 pipes, two pumps, three tanks, and two reservoirs. Net2 was excluded from pump-outage experiments because its benchmark configuration has no pumps. EPANET-compatible models were executed using the WNTR simulator with pressure-dependent demand (PDD), a 48-hour simulation window, and 900-second hydraulic/reporting timesteps. The modeled required pressure threshold was 20 m. All analyses refer to simulation outputs rather than observed utility operations.

\subsection{Study matrix and scenario semantics}\label{study-matrix-and-scenario-semantics}

The experiment is a deterministic scenario evaluation rather than an observational randomized trial. Hydraulic configurations modify benchmark input conditions to explore how response sensitivity changes under controlled perturbations. Six configurations are retained exactly as encoded in the v5.5 plan; they are not ordered as empirically calibrated drought severities. The interventions are hypothetical response capabilities: S1 changes demand by a specified percentage, S2 reduces imposed outage duration, and S3 combines these mechanisms. All interpretations must be conditional on those intervention assumptions. The attack labels A1 and A2 refer to pump-control disruptions (not verified network intrusions), while A0 denotes the no-disruption comparison. No field deployment, remote-control access, or attacker success-rate model is part of the design.

\subsection{Factorial scenario design}\label{factorial-scenario-design}

Six hydraulic configurations were evaluated: baseline (demand factor 1.00; reservoir-head shift 0 m), demand +3\%, demand +10\%, source-head -0.5 m, source-head -2 m, and combined demand +10\% with source-head -2 m. These numeric perturbations were selected for controlled hydraulic sensitivity analysis, not inferred from the U.S. Drought Monitor or streamflow records. Each configuration was crossed with attack A0 (none), A1 (four-hour pump outage), and A2 (twelve-hour pump outage); strategies S0 (none), S1 (8\% demand reduction), S2 (attack duration multiplied by 0.5), and S3 (both); and attack start times of 8 and 16 hours. The resulting file held 288 records, including 48 cached repetitions and 240 distinct configuration keys. Neither cached repetitions nor multiple deterministic scenarios are assumed to constitute independent field replications.

\subsection{Intervention validation}\label{intervention-validation}

Early simulator versions exposed two implementation errors: modifications to control conditions could alter Net1 tank state before the scheduled attack, and priority-only pump closure could be superseded by the ordinary low-tank pump-start rule. Net1 ordinarily starts pump 9 when tank 2 level falls below 33.528 m and stops it when the level exceeds 42.672 m. A final time-windowed pump-start restriction preserved normal behavior outside the imposed outage and suppressed the competing pump-start action inside it. This is a simulated control intervention, not a real-world compromise procedure.

Each attacked simulation was paired with the same network, hydraulic stress, strategy, start time, and pump selection without the attack. Validation required (i) matching pressure and tank-head trajectories before attack to within 0.01 m, (ii) matching pre-attack pump states, and (iii) closed pump status and near-zero recorded flow throughout the scheduled outage. All 192 attack comparisons passed checks at the 15-minute reported timesteps. This sampled verification does not rule out unobserved sub-timestep control events.

\subsection{Counterfactual matching, repeated configurations, and numerical validity}\label{counterfactual-matching-repeated-configurations-and-numerical-validity}

For an attacked run indexed by network \(n\), stress configuration \(s\), intervention \(m\), start time \(t_0\), and target pump \(p\), the reference is the A0 run with identical indices except for attack type. A pair is valid only if reporting-time pressures, tank head, and selected-pump states match before the scheduled outage; this provides a stronger check than merely completing the simulations without software exceptions. Hydraulic simulation records are sampled every 900 seconds, and audit claims should therefore be confined to those observed instants. No exploitability, continuous-time control invariant, or field equipment property is inferred.

The 288 rows include 48 repeated cached configurations (240 unique simulation keys). Therefore, the unique scenarios, not the total saved file rows, define independent \emph{configurations}; even these are not independent statistical realizations of water utilities. The 192 attack comparisons have distinct simulation keys and a paired audit, but outcomes share common network models and deterministic design factors. Accordingly, descriptive contrasts and controlled sensitivity results are more defensible than broad population-based confidence intervals or p-values derived from treating simulations as independent field samples.

\subsection{Outcome measures}\label{outcome-measures}

Requested demand volume was integrated from reference demand patterns; delivered demand volume was computed from PDD simulator outputs. Unmet demand is

\[U=\sum_{t}\sum_i \max(0,D^{req}_{i,t}-D^{del}_{i,t})\Delta t.\]

Water service availability is \(WSA=1-U/V_{req}\) for positive requested volume \(V_{req}\). Normalized pressure deficit averages \(\max(0,20-P_{i,t})/20\) across junctions and the simulated time intervals. Low-pressure node-hours accumulate junction-time spent below 20 m. The minimum tank level is retained as a physical-state diagnostic.

For a loss outcome \(Y\), the additive interaction contrast for a given stress level \(s\) and attack \(a\) is:

\[I(s,a)=[Y(s,a)-Y(s,0)]-[Y(0,a)-Y(0,0)].\]

Positive values indicate superadditivity on that outcome scale; negative values indicate subadditivity. Averaging across deterministic configurations describes the selected scenario grid rather than a sampled population.

\subsection{Recovery and relapse definitions}\label{recovery-and-relapse-definitions}

For matched attacked and no-attack trajectories, maximum junction pressure deviation at time \(t\) is \(\Delta_P(t)=\max_i|P_{i,a}(t)-P_{i,0}(t)|\). The pressure recovery tolerance was \(\epsilon_P=0.5\) m. \emph{First recovery} requires a one-hour post-outage window in which \(\Delta_P(t)\leq\epsilon_P\); \emph{sustained recovery} additionally requires continued compliance to the end of observation. A relapse is any subsequent excursion outside tolerance after initial qualifying recovery. Cumulative relapse duration sums out-of-tolerance reporting intervals occurring after first recovery. Results that do not satisfy a defined endpoint before the simulation ends are right-censored. Water-service recovery is assessed separately by comparing the attacked and no-attack unmet-demand-rate trajectories under the package's predefined numerical criterion. A zero post-attack residual recovery time does not imply an attack caused no during-attack harm.

\subsection{Analysis and limits of inference}\label{analysis-and-limits-of-inference}

The results are descriptive: two benchmark networks and deterministic perturbations do not support extrapolating event frequencies or confidence intervals to real utilities. Strategy comparisons use absolute outcomes and recovery counts, not solely interaction contrasts. No measured drought-to-hydraulic mapping was used in the final v5.5 scenario records.

\subsection{Recovery algorithm and censoring}\label{recovery-algorithm-and-censoring}

Censoring is retained when a qualifying pressure-recovery event is not demonstrated within the 48-hour observation horizon. This time-to-event framing is related to the treatment of incomplete observation windows in Kaplan--Meier estimation \cite{kaplan}; we do not report Kaplan--Meier survival curves for these deterministic scenario counts.

Let \(P_j^{A}(t)\) and \(P_j^{0}(t)\) denote attacked and paired-reference pressure at junction \(j\). At each recorded post-outage time, compare the maximum absolute deviation \(D_P(t)=\max_j |P_j^A(t)-P_j^0(t)|\) to the preset 0.5 m threshold. The first recovery is the first starting sample of a complete one-hour compliant window. Sustained recovery requires a qualifying window after whose start no subsequent sampled pressure deviation exceeds tolerance before the simulation ends. A relapse episode is a new contiguous interval of exceedance following the first qualifying recovery window. End-state compliance and episode duration are reported separately. If the condition is never met by hour 48, the event is right-censored rather than coded as a zero-hour recovery time.

Water-service recovery follows the analogous matched unmet-demand-rate criterion (tolerance set by the v5.6 analysis implementation). End-of-horizon sustained recovery is only sustained \emph{within the observed period}, not proof of indefinite convergence. Results should be checked under alternative pressure tolerances and horizons before extrapolation.

\subsection{Observational environmental data and statistical analysis}\label{observational-environmental-data-and-statistical-analysis}

The regional environmental analysis uses 313 weekly county-level Drought Severity and Coverage Index (DSCI) observations (Ingham County, FIPS 26065), 2,192 daily USGS discharge observations at Red Cedar River station 04112500, and 2,192 daily NASA POWER temperature/precipitation observations for January 2020 through December 2025 \cite{usgs,usdm,nasa}. Daily inputs were aligned to the weekly USDM reporting date. The DSCI is a composite drought severity/coverage indicator on a 0--500 scale rather than a direct water availability or reservoir head measurement \cite{usdm}. Same-week and lagged Spearman correlations were calculated. For the regression, weekly mean river discharge was log-transformed, and additive linear models included seasonal Fourier components, temperature, same-week precipitation, and (in the fullest model) four-week antecedent precipitation. Heteroskedasticity- and autocorrelation-consistent (HAC) uncertainty estimates were computed following the logic of Newey--West covariance estimation \cite{newey}. Classical nonparametric trend and robust-slope estimators \cite{mann,sen} provide useful sensitivity-analysis alternatives, although they were not applied in the present reported regressions. The observational association is not interpreted as a causal drought effect, nor is the river gauge assumed to supply Net1, Net3, or Lansing's municipal distribution system.

\section{Results}\label{results}

\subsection{Simulation completion and intervention verification}\label{simulation-completion-and-intervention-verification}

All 288 records completed successfully. The file contains 192 attack records and 96 no-attack records. All 192 attack scenarios passed pump-outage verification, matched pre-attack pressure-equivalence checks, tank-head equivalence, and pre-attack pump-status equivalence. The full audit reported maximum pre-attack junction-pressure discrepancies on the order of 10\^{}-13 m, substantially below the 0.01 m criterion. Forty-eight rows represented cached repetitions; thus, 240 distinct simulation keys were present.

\subsection{Network-dependent hydraulic losses}\label{network-dependent-hydraulic-losses}

Table~\ref{tab:results1} summarizes the primary service and pressure outcomes.

\begin{longtable}[]{@{}
  >{\raggedright\arraybackslash}p{0.1517\linewidth}
  >{\raggedright\arraybackslash}p{0.1517\linewidth}
  >{\raggedleft\arraybackslash}p{0.2022\linewidth}
  >{\raggedleft\arraybackslash}p{0.2022\linewidth}
  >{\raggedleft\arraybackslash}p{0.2022\linewidth}@{}}
\caption{Network-level hydraulic outcomes across attack configurations}\label{tab:results1}\\
\toprule\noalign{}
\begin{minipage}[b]{\linewidth}\raggedright
Network
\end{minipage} & \begin{minipage}[b]{\linewidth}\raggedright
Attack
\end{minipage} & \begin{minipage}[b]{\linewidth}\raggedleft
Mean unmet demand (m$^3$)
\end{minipage} & \begin{minipage}[b]{\linewidth}\raggedleft
Mean normalized pressure deficit
\end{minipage} & \begin{minipage}[b]{\linewidth}\raggedleft
Mean WSA
\end{minipage} \\
\midrule\noalign{}
\endhead
\bottomrule\noalign{}
\endlastfoot
Net1 & A0 & 0 & 0 & 1.000000 \\
Net1 & A1, 4 h & 0 & 0 & 1.000000 \\
Net1 & A2, 12 h & 404.96 & 0.043511 & 0.966276 \\
Net3 & A0 & 0 & 0.023894 & 1.000000 \\
Net3 & A1, 4 h & 0 & 0.024765 & 1.000000 \\
Net3 & A2, 12 h & 0 & 0.025593 & 1.000000 \\
\end{longtable}

Longer outages in Net1 generated measurable unmet demand and pressure deficits. Net3 retained full modeled demand delivery over the evaluated configurations but exhibited modest deterioration in its pressure-deficit metric. This difference is consistent with distinct storage/control and hydraulic characteristics but does not isolate any single causal network feature.

\subsection{Strategy sensitivity under Net1 twelve-hour attacks}\label{strategy-sensitivity-under-net1-twelve-hour-attacks}

The intervention contrasts are summarized in Table~\ref{tab:results2}.

\begin{longtable}[]{@{}
  >{\raggedright\arraybackslash}p{0.1820\linewidth}
  >{\raggedleft\arraybackslash}p{0.2427\linewidth}
  >{\raggedleft\arraybackslash}p{0.2427\linewidth}
  >{\raggedleft\arraybackslash}p{0.2427\linewidth}@{}}
\caption{Intervention sensitivity for Net1 twelve-hour outages}\label{tab:results2}\\
\toprule\noalign{}
\begin{minipage}[b]{\linewidth}\raggedright
Strategy
\end{minipage} & \begin{minipage}[b]{\linewidth}\raggedleft
Mean unmet demand (m$^3$)
\end{minipage} & \begin{minipage}[b]{\linewidth}\raggedleft
Mean normalized pressure deficit
\end{minipage} & \begin{minipage}[b]{\linewidth}\raggedleft
Mean WSA
\end{minipage} \\
\midrule\noalign{}
\endhead
\bottomrule\noalign{}
\endlastfoot
S0: no intervention & 819.05 & 0.086806 & 0.934288 \\
S1: demand reduction & 732.44 & 0.082031 & 0.936024 \\
S2: shortened outage & 68.34 & 0.005208 & 0.994792 \\
S3: combined & 0.00 & 0.000000 & 1.000000 \\
\end{longtable}

Within the simulation grid, S3 eliminated unmet demand in Net1 A2 scenarios, whereas S2 alone markedly reduced mean losses. The assumed strategy effects---8\% demand reduction and a 50\% reduction in imposed outage duration---are experimental settings rather than measured capacities of a deployed intervention. Comparisons also require care because demand management changes the amount of requested water and thus the WSA denominator.

\subsection{Matched compound-stress interactions}\label{matched-compound-stress-interactions}

Table~\ref{tab:results3} reports the matched interaction summaries.

\begin{longtable}[]{@{}
  >{\raggedright\arraybackslash}p{0.1950\linewidth}
  >{\raggedright\arraybackslash}p{0.1950\linewidth}
  >{\raggedleft\arraybackslash}p{0.2600\linewidth}
  >{\raggedleft\arraybackslash}p{0.2600\linewidth}@{}}
\caption{Matched compound-stress interaction summaries}\label{tab:results3}\\
\toprule\noalign{}
\begin{minipage}[b]{\linewidth}\raggedright
Network
\end{minipage} & \begin{minipage}[b]{\linewidth}\raggedright
Attack
\end{minipage} & \begin{minipage}[b]{\linewidth}\raggedleft
Mean pressure-deficit interaction
\end{minipage} & \begin{minipage}[b]{\linewidth}\raggedleft
Mean unmet-demand interaction (m$^3$)
\end{minipage} \\
\midrule\noalign{}
\endhead
\bottomrule\noalign{}
\endlastfoot
Net1 & A1 & 0 & 0 \\
Net1 & A2 & +0.002214 & +34.97 \\
Net3 & A1 & +0.000023 & 0 \\
Net3 & A2 & +0.000010 & 0 \\
\end{longtable}

For Net1 A2, unmet-demand interaction contrasts ranged from approximately -526.36 to +602.11 m$^3$. Therefore, the modest positive mean does not support a blanket statement that hydraulic stress always amplifies cyber-induced disruption. The interaction is conditional on network, operating assumptions, strategy, and attack timing.

\subsection{First and sustained pressure recovery}\label{first-and-sustained-pressure-recovery}

The network- and duration-specific recovery contrasts appear in Table~\ref{tab:results4}.

All 192 attack trajectories passed the matched pre-attack validation. First pressure recovery occurred in 129 records and sustained pressure recovery in 103. Relapse occurred at least once in 65 records; the aggregate relapse duration was 197.75 scenario-hours.

\begin{longtable}[]{@{}
  >{\raggedright\arraybackslash}p{0.1517\linewidth}
  >{\raggedright\arraybackslash}p{0.1517\linewidth}
  >{\raggedleft\arraybackslash}p{0.2022\linewidth}
  >{\raggedleft\arraybackslash}p{0.2022\linewidth}
  >{\raggedleft\arraybackslash}p{0.2022\linewidth}@{}}
\caption{First and sustained pressure recovery by network and attack}\label{tab:results4}\\
\toprule\noalign{}
\begin{minipage}[b]{\linewidth}\raggedright
Network
\end{minipage} & \begin{minipage}[b]{\linewidth}\raggedright
Attack
\end{minipage} & \begin{minipage}[b]{\linewidth}\raggedleft
First pressure recovery
\end{minipage} & \begin{minipage}[b]{\linewidth}\raggedleft
Sustained pressure recovery
\end{minipage} & \begin{minipage}[b]{\linewidth}\raggedleft
Cases with relapse
\end{minipage} \\
\midrule\noalign{}
\endhead
\bottomrule\noalign{}
\endlastfoot
Net1 & A1 & 24/48 & 20/48 & 18/48 \\
Net1 & A2 & 11/48 & 10/48 & 6/48 \\
Net3 & A1 & 48/48 & 38/48 & 18/48 \\
Net3 & A2 & 46/48 & 35/48 & 23/48 \\
\end{longtable}

Relapse and sustained recovery are not mutually exclusive: a case can relapse after first recovery and subsequently recover sustainably. Seventy-four cases ended outside the 0.5 m pressure-trajectory tolerance. Censored trajectories must not be assigned zero or arbitrary complete-recovery times.

\subsection{Recovery by intervention and service metric}\label{recovery-by-intervention-and-service-metric}

Table~\ref{tab:results5} compares intervention-specific recovery outcomes.

\begin{longtable}[]{@{}
  >{\raggedright\arraybackslash}p{0.1820\linewidth}
  >{\raggedleft\arraybackslash}p{0.2427\linewidth}
  >{\raggedleft\arraybackslash}p{0.2427\linewidth}
  >{\raggedleft\arraybackslash}p{0.2427\linewidth}@{}}
\caption{Recovery outcomes by intervention strategy}\label{tab:results5}\\
\toprule\noalign{}
\begin{minipage}[b]{\linewidth}\raggedright
Strategy
\end{minipage} & \begin{minipage}[b]{\linewidth}\raggedleft
First pressure recovery
\end{minipage} & \begin{minipage}[b]{\linewidth}\raggedleft
Sustained pressure recovery
\end{minipage} & \begin{minipage}[b]{\linewidth}\raggedleft
Cases with relapse
\end{minipage} \\
\midrule\noalign{}
\endhead
\bottomrule\noalign{}
\endlastfoot
S0 & 30/48 & 28/48 & 14/48 \\
S1 & 31/48 & 19/48 & 19/48 \\
S2 & 36/48 & 34/48 & 16/48 \\
S3 & 32/48 & 22/48 & 16/48 \\
\end{longtable}

Outage shortening (S2) produced the largest sustained pressure-recovery count (34/48), whereas the combined strategy S3 performed best on Net1 unmet demand under A2. All 192 attack records achieved sustained post-outage water-service recovery under the specified unmet-demand-rate criterion, and no service relapses were recorded. Nevertheless, 27 Net1 A2 records exhibited during-attack water-service deviation. A network can therefore resume modeled demand delivery without matching its no-attack pressure trajectory.

\subsection{Figure set derived directly from archived aggregate outputs}\label{figure-set-derived-directly-from-archived-aggregate-outputs}

\begin{figure}
\centering
\includegraphics[width=0.92\linewidth,keepaspectratio]{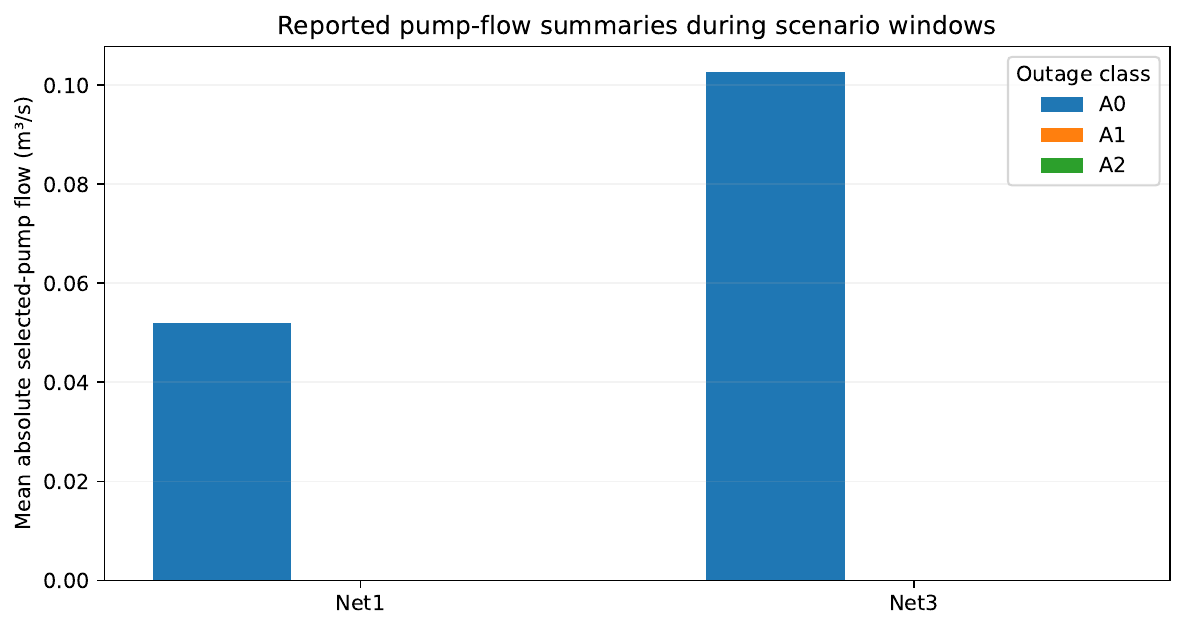}
\caption{Selected-pump flow summaries}\label{fig:result1}
\end{figure}

\begin{figure}
\centering
\includegraphics[width=0.92\linewidth,keepaspectratio]{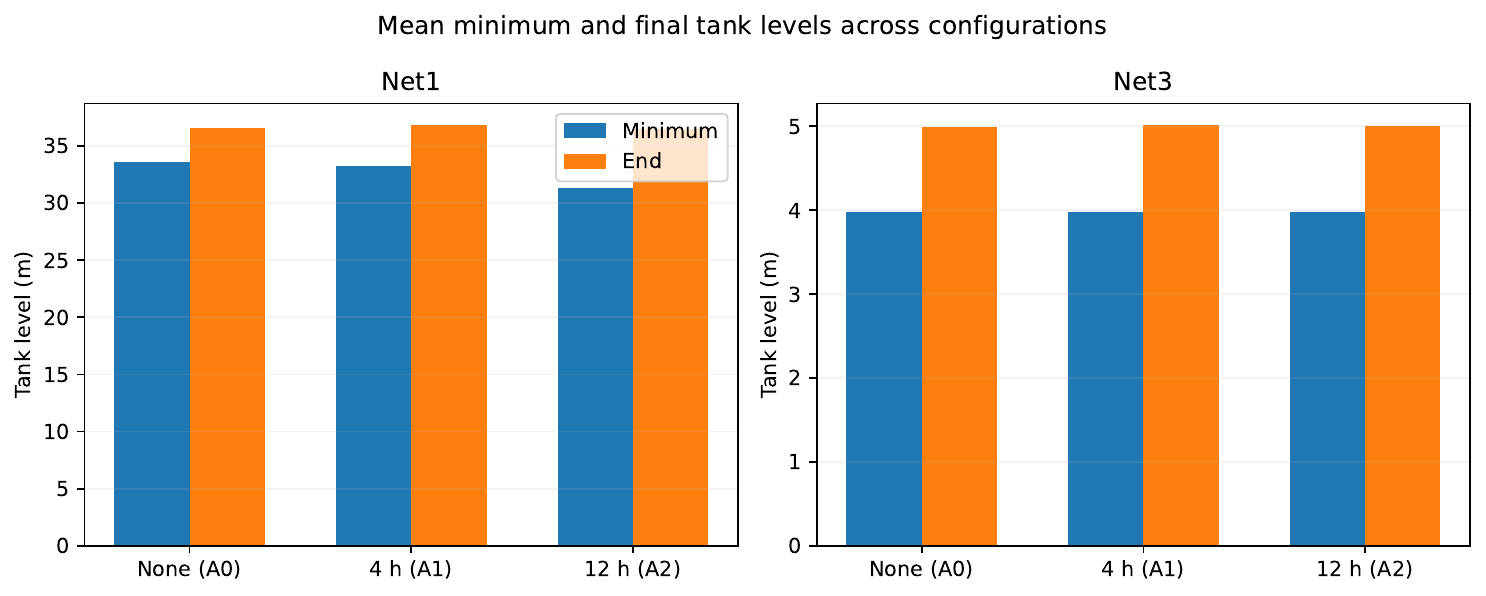}
\caption{Tank minimum and final levels}\label{fig:result2}
\end{figure}

\begin{figure}
\centering
\includegraphics[width=0.92\linewidth,keepaspectratio]{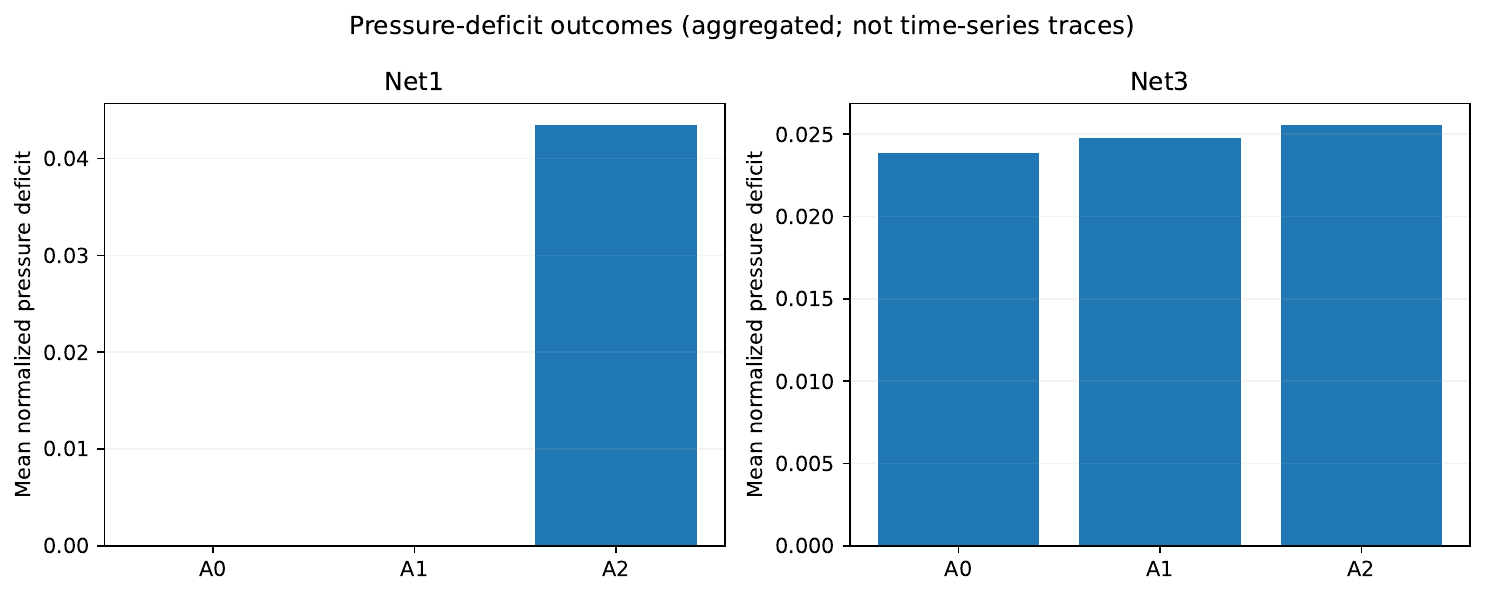}
\caption{Mean normalized pressure deficits}\label{fig:result3}
\end{figure}

\begin{figure}
\centering
\includegraphics[width=0.92\linewidth,keepaspectratio]{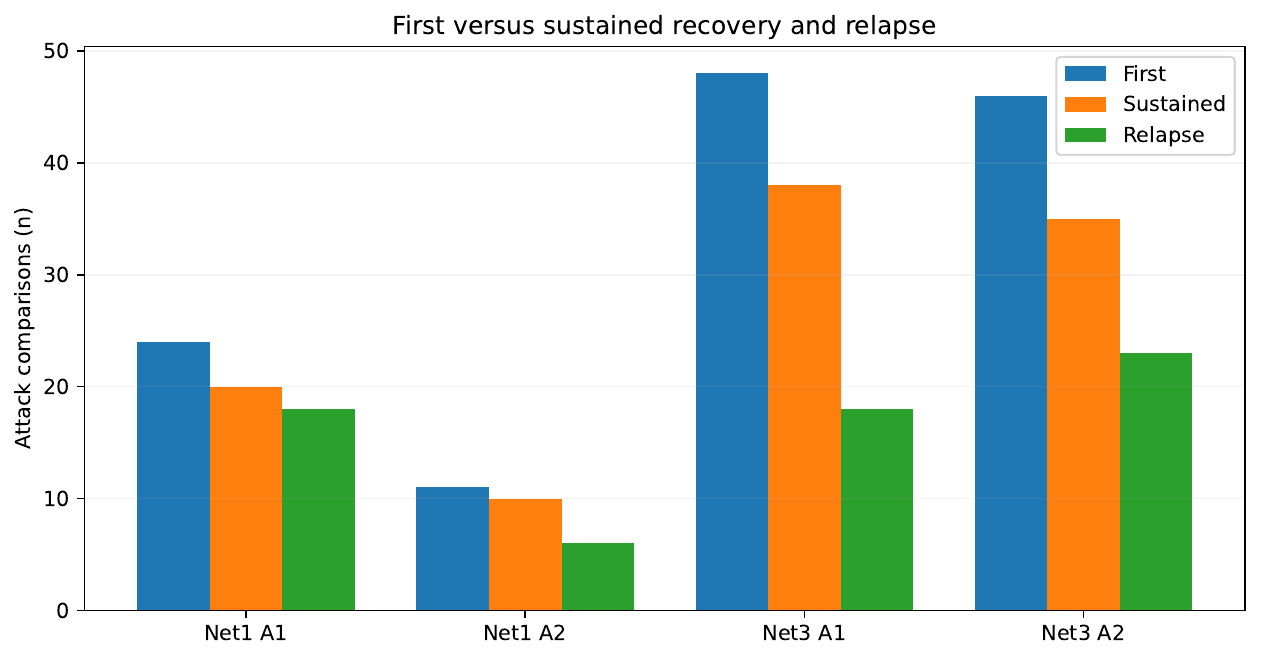}
\caption{Figure 4: first recovery, sustained recovery, and relapse}\label{fig:result4}
\end{figure}

\begin{figure}
\centering
\includegraphics[width=0.92\linewidth,keepaspectratio]{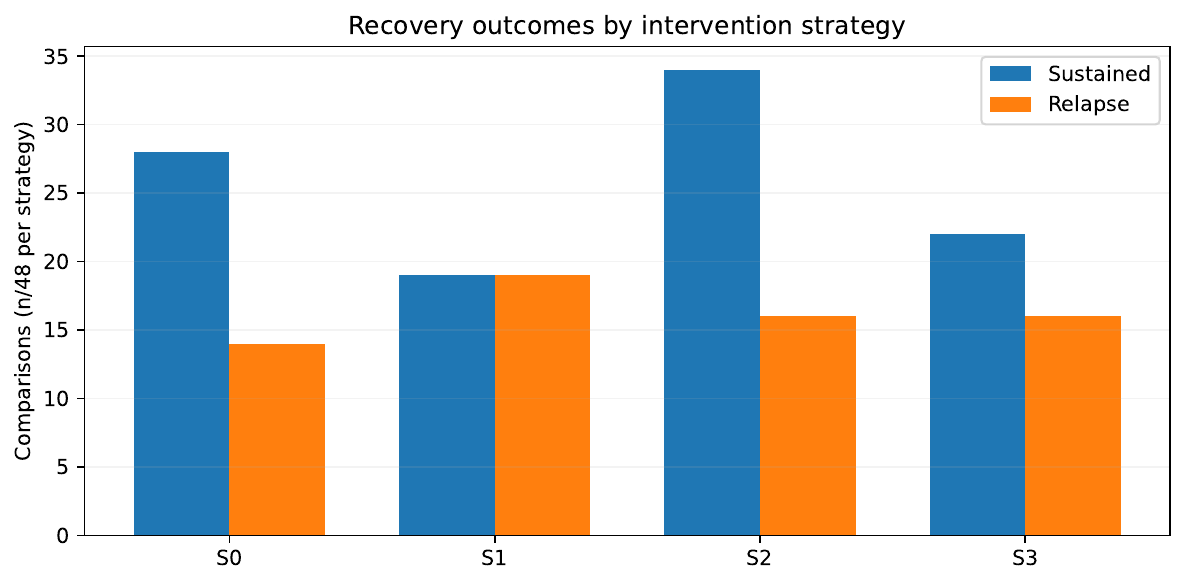}
\caption{Figure 5: recovery comparisons by intervention}\label{fig:result5}
\end{figure}

The pump-flow, tank-state, and pressure-deficit summaries appear in Figs.~\ref{fig:result1}--\ref{fig:result3}, whereas recovery/relapse and strategy comparisons appear in Figs.~\ref{fig:result4} and \ref{fig:result5}. These are aggregate figures, not time-resolved hydraulic traces.

Mean absolute selected-pump flow for each network/attack configuration, aggregated from the reported scenario windows. This is \textbf{not} a time-resolved pump-flow trace. A0 window summaries are plotted for comparison and do not represent a forced-closure verification condition.

Average minimum and final tank levels, grouped by network and attack class; the figure does not establish the exact timing of tank depletion. For multiple tanks, the stored minimum statistic follows the simulation's configured metric and is not necessarily the same tank across networks.

Aggregate normalized pressure-deficit results; longitudinal pressure-deviation curves require the original \texttt{.npz} snapshots.

Counts from the v5.6 detailed file. Recovery and relapse categories can overlap; they must not be read as partitions of a common total.

Strategy counts, 48 attack comparisons per strategy, conditioned on the chosen hydraulic configurations and benchmark networks.

\subsection{Observed drought--streamflow relationship}\label{observed-droughtstreamflow-relationship}

\begin{figure}
\centering
\includegraphics[width=0.92\linewidth,keepaspectratio]{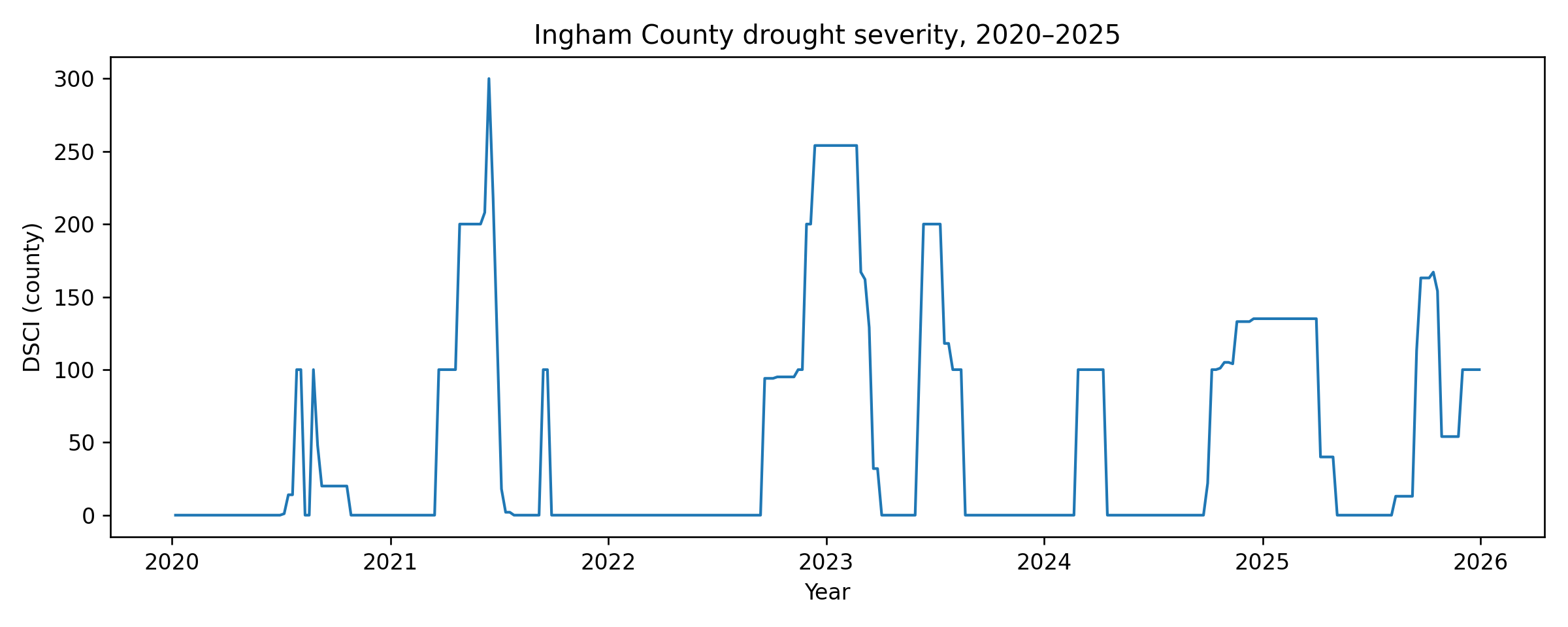}
\caption{Observed USDM drought severity in Ingham County, 2020--2025.}\label{fig:result6}
\end{figure}

\begin{figure}
\centering
\includegraphics[width=0.92\linewidth,keepaspectratio]{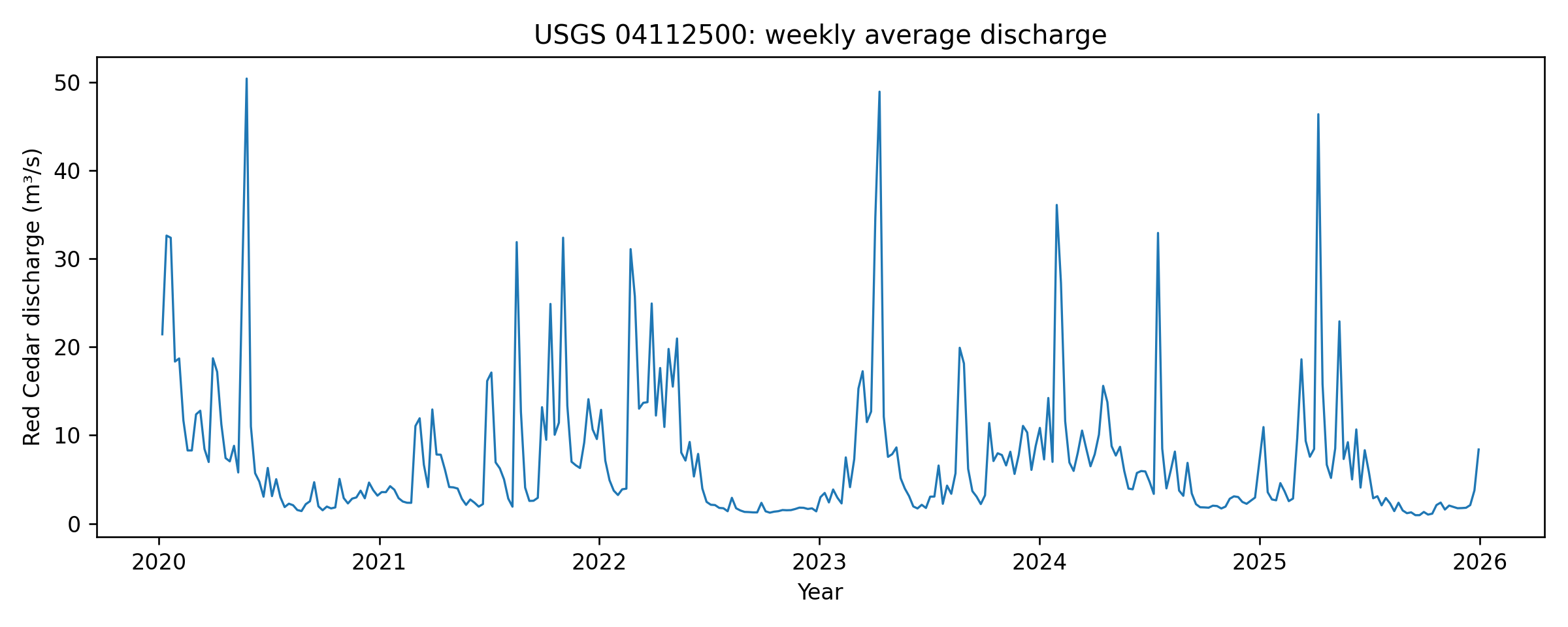}
\caption{Observed daily discharge in Red Cedar River at East Lansing, Michigan, 2020--2025.}\label{fig:result7}
\end{figure}

\begin{figure}
\centering
\includegraphics[width=0.92\linewidth,keepaspectratio]{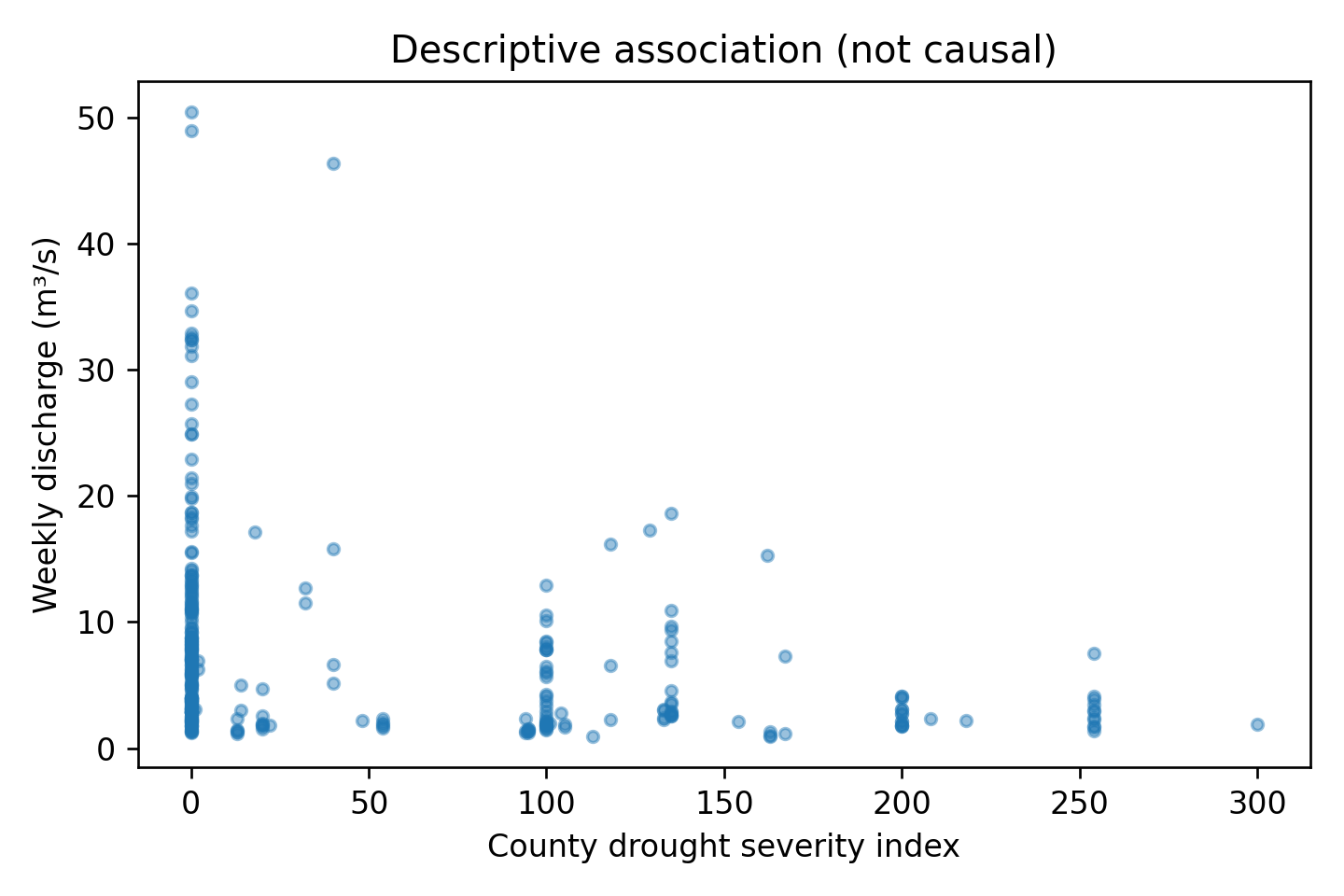}
\caption{Observed county drought--river discharge relationship; plotted observations are not municipal hydraulic measurements.}\label{fig:result8}
\end{figure}

The temporal drought, streamflow, and paired drought--discharge visualizations are shown in Figs.~\ref{fig:result6}--\ref{fig:result8}.

The climate regression estimates are provided in Table~\ref{tab:results6}.

Across 313 weekly observations, the Spearman correlation between DSCI and same-week river discharge was \(\rho=-0.4098\); the corresponding one-, two-, and four-week lagged DSCI correlations were \(-0.3062\), \(-0.2356\), and \(-0.1344\). Same-week precipitation and river discharge had a Spearman correlation of \(+0.2433\). These relationships are descriptive and temporally autocorrelated; nominal rank-correlation significance is not treated as independent-observation inference.

\begin{longtable}[]{@{}
  >{\raggedright\arraybackslash}p{0.1820\linewidth}
  >{\raggedleft\arraybackslash}p{0.1820\linewidth}
  >{\raggedleft\arraybackslash}p{0.1820\linewidth}
  >{\raggedright\arraybackslash}p{0.1820\linewidth}
  >{\raggedleft\arraybackslash}p{0.1820\linewidth}@{}}
\caption{Seasonally adjusted drought--streamflow regression models}\label{tab:results6}\\
\toprule\noalign{}
\begin{minipage}[b]{\linewidth}\raggedright
Regression specification
\end{minipage} & \begin{minipage}[b]{\linewidth}\raggedleft
Weeks
\end{minipage} & \begin{minipage}[b]{\linewidth}\raggedleft
DSCI coefficient per 100 points, log discharge
\end{minipage} & \begin{minipage}[b]{\linewidth}\raggedright
95\% HAC interval
\end{minipage} & \begin{minipage}[b]{\linewidth}\raggedleft
Adjusted R$^2$
\end{minipage} \\
\midrule\noalign{}
\endhead
\bottomrule\noalign{}
\endlastfoot
Seasonal Fourier terms & 313 & -0.4820 & {[}-0.6047, -0.3592{]} & 0.436 \\
Seasonal + current weather & 313 & -0.4848 & {[}-0.5918, -0.3777{]} & 0.555 \\
Seasonal + weather + prior four-week precipitation & 309 & -0.3246 & {[}-0.4234, -0.2258{]} & 0.661 \\
\end{longtable}

In the fullest model, the fitted contrast for a 100-point higher DSCI corresponds to \(\exp(-0.3246)-1\approx -27.7\%\) in conditional weekly mean discharge. This exploratory association is sensitive to model assumptions and is not a physical conversion between flow and pressure. It also does not provide a detected long-term climate-change trend; only six observational years are used.

\section{Discussion}\label{discussion}

\subsection{Control validity is part of experimental validity}\label{control-validity-is-part-of-experimental-validity}

The development audit demonstrated that nominal simulation completion is not sufficient to infer a correctly imposed outage. In Net1, unmodified low-tank pump-start logic could reopen the pump during a long disruption, while an overly broad gating method could shift tank head before the intended start time. A time-windowed restriction achieved both counterfactual equivalence and reported pump closure in the final tests. This finding is methodological, not evidence that the benchmark network exposes an exploitable vulnerability. It demonstrates that physical control logic must be audited when operational disturbances are used as security proxies.

\subsection{Service continuity and hydraulic-state restoration diverge}\label{service-continuity-and-hydraulic-state-restoration-diverge}

All attacks met the post-outage service criterion, yet only 103/192 achieved sustained pressure recovery. Tank storage, pump-control hysteresis, and changing source/demand configurations can keep a network away from its counterfactual pressure trajectory even after requested demand is again served. Thus, a single metric such as WSA is insufficient for longitudinal resilience assessment. Recovery must be defined against a suitable baseline, over a stated tolerance and observation horizon, with relapses and right-censoring reported explicitly.

\subsection{Strategy selection is outcome-dependent}\label{strategy-selection-is-outcome-dependent}

S3 minimized Net1 A2 unmet demand in the evaluated scenarios, whereas S2 yielded the highest count of sustained pressure recovery across both networks. This is not evidence that demand reduction undermines actual resilience: changing demands also changes pressures, reference trajectories, and the physical state being restored. Instead, it shows that ranking interventions without declaring the resilience endpoint can be misleading. Utilities may reasonably prioritize different outcomes depending on critical customers, minimum acceptable pressure, storage risk, and operational costs.

\subsection{Drought and climate adaptation: a two-stage evidence chain}\label{drought-and-climate-adaptation-a-two-stage-evidence-chain}

The climate component requires two distinct links. The first is \emph{environmental association}: do weekly area-based USDM severity measurements covary with locally relevant river flow, groundwater levels, meteorological drought indices, or reservoir storage? The second is \emph{utility hydraulic translation}: do those environmental conditions predict or constrain available source head, delivered production, demand, tank storage, or pump operating modes at the actual facility? While the first can potentially be estimated from public observations, the second is utility- and infrastructure-specific. It cannot be inferred merely because a gauge lies in the same region or because an EPANET parameter can be reduced.

A future calibrated analysis should first verify watershed/intake relevance and source type. For example, a utility supplied primarily by groundwater would not generally justify using Red Cedar River discharge to set hydraulic reservoir head without a demonstrated aquifer connection. Then, measured operational data could support competing parameterizations, out-of-sample tests, and uncertainty intervals. The finalized cyber--hydraulic controls could be reused without modifying attack implementation. This staged design avoids confounding physically valid cyber-disruption results with unsupported climate calibration.

\subsection{Practical meaning for operators and policymakers}\label{practical-meaning-for-operators-and-policymakers}

Operational response planning may emphasize immediate customer delivery, a minimum pressure floor, secure pump restart, or long-term storage recovery. The results show why choosing a single resilience endpoint is consequential: the combined action removes modeled Net1 A2 unmet-demand loss, whereas shorter outage duration produces more persistent counterfactual pressure matches in the scenario matrix. Neither is guaranteed to be optimal for utilities with different tank geometries, seasonal operating rules, energy tariffs, or critical-customer constraints. From the standpoint of climate adaptation policy, the reliable near-term contribution is an auditable evaluation \emph{method}: models should verify planned cyber interventions, retain a physically consistent baseline, and monitor post-event relapse rather than equating reopening an actuator with full resilience.

\subsection{Environmental and climate adaptation relevance}\label{environmental-and-climate-adaptation-relevance}

Water-demand and source-head stress tests are relevant to assessing operating margins under scarcity, seasonal variability, and other possible stressors. However, no observed USGS discharge series, drought index, or historical utility flow record was used to calibrate the final simulated demand multipliers or reservoir-head reductions. Accordingly, the experiment cannot establish drought causation, climate-change attribution, or utility-level adaptation effectiveness. A future study could use a calibrated demand model, utility production data, storage operations, and physically supported surface-water/groundwater linkage to establish such relationships. This step is necessary before describing stress interactions as empirical cyber--climate effects.

\subsection{Robustness checks needed for external validity}\label{robustness-checks-needed-for-external-validity}

The present evidence supports several specific falsification tests for the next stage: rerun recovery definitions at 0.25 m, 0.5 m, and 1.0 m tolerances; evaluate one- versus two-hour qualifying intervals; extend the observation horizon beyond 48 hours; test controls at a finer hydraulic reporting interval; and repeat the attack design on a third network with a different supply architecture. These changes should be treated as prospective additional analyses, not as tests that have already passed. The observational climate model should use season-aware lag structures, benchmark against season-only models, and avoid selecting only convenient extreme historical dates. None of these tests should be interpreted as achieved until executed and archived.

\section{Limitations and threats to validity}\label{limitations-and-threats-to-validity}

The findings derive from two small-to-medium benchmark networks rather than independently calibrated utilities. Pump outages stand in for operational effects potentially induced by a cyber event; no intrusion vector, adversarial communication process, or actual attack likelihood was simulated. Outage checks were sampled at 15-minute reporting intervals, and unobserved internal solver events remain possible. The environmental perturbations are hypothetical sensitivity settings; historical climatic observations do not identify counterfactual hydraulic states. There are 48 cached duplicate records, and deterministic sweeps over chosen settings are not independent statistical replications. The 0.5 m recovery tolerance and one-hour qualification window are operational definitions; robustness to alternative tolerances and observation horizons should be explored. Recovery observations are bounded at 48 hours, so censoring is material. Finally, the outcome set does not address water quality, contamination, energy cost, equity, or actual emergency response capacity.

\section{Conclusions}\label{conclusions}

A paired, validated WNTR experiment on Net1 and Net3 demonstrates that the consequences of simulated pump-control interruptions depend on hydraulic configuration and on the endpoint used to assess resilience. In the 288-record experiment, all 192 scheduled attack cases passed reported outage and pre-attack equivalence audits. Twelve-hour outages caused mean unmet demand of 404.96 m$^3$ in Net1, while Net3 preserved modeled demand delivery. Across all attacks, first pressure recovery occurred in 129 records, sustained pressure recovery in 103, and relapse in 65, even though every case met the post-outage water-service recovery criterion. No single tested strategy dominated all endpoints: combined demand reduction and outage shortening eliminated modeled Net1 A2 unmet demand, whereas outage shortening alone yielded the highest sustained pressure-recovery count. These results support a methodological recommendation to validate imposed disturbances and report both service continuity and durable physical-state recovery. Claims about climate-change effects and broader utility populations require observed-data calibration and independent-network validation.

\section{Reproducibility and release status}\label{reproducibility-and-release-status}
The research materials supporting this study have been
publicly archived on Zenodo under DOI:
\href{https://doi.org/10.5281/zenodo.23477358}
{10.5281/zenodo.23477358}.

The reproducibility package includes the available v5.5
simulation source files, the v5.6 recovery-analysis
algorithm, simulation summary and validation tables,
derived aggregate results, visualization materials,
dependency specifications, a checksum manifest, and
scripts for analyzing the observed drought--streamflow
relationship.

The archive supports examination of the experimental
methodology, verification of reported aggregate findings,
and independent extension of the proposed cyber--hydraulic
resilience framework. However, the original time-resolved
hydraulic trajectory files (\texttt{.npz}) and independently
verified benchmark input files were not included in the
materials available when the package was assembled.
Consequently, complete reproduction of every original
simulation trajectory cannot be guaranteed from the
archived materials alone.

Furthermore, the specified dependency versions represent
a proposed computational environment rather than a
verified, byte-identical reconstruction of the original
experimental environment. A clean-environment execution,
complete input provenance verification, and dependency
consistency checks remain necessary to establish full
computational reproducibility.

The public research archive is permanently accessible at
\url{https://doi.org/10.5281/zenodo.23477358}.

\section{Data availability, ethics and authorship}\label{data-availability-ethics-and-authorship}

The analysis is based on the v5.5 simulation results
(\texttt{results\_v55.csv}), paired-control validation
records, and v5.6 detailed and summary recovery datasets.
The available simulation code, experimental results,
validation records, and supporting analysis materials
have been publicly archived on Zenodo and are accessible
at \url{https://doi.org/10.5281/zenodo.23477358}.

The archived materials support transparency, reproducibility
assessment, and independent evaluation of the proposed
cyber--hydraulic resilience framework. Nevertheless,
complete computational reproduction remains subject to
the availability of original time-resolved simulation
trajectories, verification of the computational environment
and software dependencies, and confirmation of benchmark
network provenance and licensing.

All cyber-induced pump disruptions were conducted
exclusively within offline hydraulic simulation environments.
No operational water distribution infrastructure, production
control system, or live utility network was accessed.

\end{document}